\documentclass[15pt,thmsa,12pt,a4paper,draft,onecolumn]{article}
\usepackage{sw20lart}

\input tcilatex
\QQQ{Language}{
American English
}

\begin{document}

\title{Effects on the hadron propagators due to $k_\mu k_\nu $ terms in the vector
meson propagator}
\author{H. X. Zhang and S. S. Wu \\
Center for Theoretical Physics, College of Physics,\\
Jilin University, Changchun 130023, P. R. China}
\date{}
\maketitle

\begin{abstract}
In an approximation where the baryon current conservation is violated, the
contribution of the $k_\mu k_\nu $ terms in the vector meson propagator may
not vanish. Their effects on the baryon and meson spectral functions and on
the consequences of self-consistency are studied in the relativistic
self-consistent Hartree-Fock approximation by means of the $\sigma -\omega $
model. Two cases where the $k_\mu k_\nu $ terms are and are not neglected
are compared. It is found that there is a marked change in the baryon
spectral function which becomes more peaked in the latter case. Such a
change remains even by a proper readjustment of parameters. The effects of
self-consistency in the $\sigma -\omega $ model are qualitatively the same
in both cases, though quantitatively there is some significant difference.
\end{abstract}

\newpage\ 

\baselineskip=1cm

\section{Introduction}

If baryons couple only with $\omega $-mesons ($\omega $ case), Krein,
Nielsen, Puff and Wilets [1] found that in the relativistic self-consistent
Hartree-Fock (RSCHF) [2, 3] calculation of the renormalized baryon
propagator, its spectral function $A_R(\kappa )$ can be negative for some
real values of $\kappa $. They emphasized that this result is unacceptable.
The spectral representation they considered is of the form 
\[
G(k)=-\int_{-\infty }^{+\infty }d\kappa A_R(\kappa )\frac{\gamma _\mu k_\mu
+i\kappa }{k^2+\kappa ^2-i\varepsilon }. 
\]
Since $A_R(\kappa )$ represents the probability that a state of mass $%
|\kappa |$ is created, it must be non-negative. They suggested that it might
be due to the inadequacy of the HF approximation or the inconsistency of the
theory. The $\omega $-meson propagator can be written as [4, 5] 
\begin{equation}
D_{\mu \nu }(k)=(\delta _{\mu \nu }-\frac{k_\mu k_\nu }{k^2})\Delta _{%
\mathrm{v}}(k)-i\frac{k_\mu k_\nu }{k^2(m_{\mathrm{v}}^2+\delta m_{\mathrm{v}%
}^2)},  \tag{1a}
\end{equation}
\begin{equation}
\Delta _{\mathrm{v}}(k)=\Delta _{\mathrm{v}}^0(k)+\Delta _{\mathrm{v}%
}^0(k)\Pi _{\mathrm{v}}(k)\Delta _{\mathrm{v}}(k)=-i[k^2+m_{\mathrm{v}%
}^2+i\Pi _{\mathrm{v}}(k)-i\varepsilon ]^{-1},  \tag{1b}
\end{equation}
where $\delta m_{\mathrm{v}}^2$ is the mass counterterm for the $\omega $%
-meson. In their calculation they have neglected all the terms proportional
to $k_\mu k_\nu $ in the $\omega $-meson propagator on the basis of the
baryon current conservation implied by the model for a rigorous calculation.
Though this is a generally accepted approximation [6] and indeed, such terms
need not be taken into account if the baryon current conserves [7], their
contribution in the RSCHF approximation is not zero and has to be studied
[8]. It also indicates that the RSCHF approximation does not preserve the
baryon current conservation.

In Ref. [8] we showed that the negative baryon spectral function mentioned
above is caused by the $k_\mu k_\nu $ terms in the $\omega $-meson
propagator. In Eq. (1a) the last longitudinal term is not renormalizable,
thus if the $k_\mu k_\nu $ terms should be considered, it must be studied
carefully. In order to take a proper account of the contribution of the $%
k_\mu k_\nu $ terms in $D_{\mu \nu }(k)$ a rule has also been proposed in
[8]. Refs. [9, 12] have shown if in addition to the $\omega $-meson, other
mesons like $\pi $, $\sigma $, and chiral $\pi -\sigma $ are considered, the
baryon spectral functions in the RSCHF approximation can be regular for
parameters of physical interest, even though the $k_\mu k_\nu $ terms in Eq.
(1a) are neglected. Clearly this does not mean that the contribution of the $%
k_\mu k_\nu $ terms is not important when the baryon current conservation is
violated, because it is related with the relative strength between different
fields. For example, in the $\sigma -\omega $ model, if we adjust the
coupling constants $g_{\mathrm{s}}^2$ and $g_{\mathrm{v}}^2$, one finds that
along with $g_{\mathrm{s}}^2/g_{\mathrm{v}}^2$ becoming smaller the
undesirable negative spectral function will appear again. So, it is
desirable to assess the effect of $k_\mu k_\nu $ terms, even in cases where
there are other mesons. The role of the $k_\mu k_\nu $ terms in the vector
meson propagator has been studied in Ref. [6]. It was found that under the $%
G_D$ approximation the effect of the $k_\mu k_\nu $ terms on the observable
quantities is negligible. In this paper we would like to consider the other
case where the baryon propagator incorporates the propagation of virtual
baryons and antibaryons. We shall study the contribution of the $k_\mu k_\nu 
$ terms to the baryon and meson spectral functions and their influence on
the effects of self-consistency by means of the rule suggested in Ref. [8].
We find that in the $\sigma -\omega $ model the regularity of the effects of
self-consistency is almost the same as found in Ref. [9], and the
contribution of the $k_\mu k_\nu $ terms to the meson spectral function is
not very important. However, there is a marked change in the baryon spectral
function which becomes more peaked. Moreover, we cannot remove such a change
by a proper readjustment of parameters. This shows that violation of the
current conservation law may cause quite significant effects.

The paper is organized as follows. In Section 2, we shall consider the
coupled set of Dyson-Schwinger (DS) equations for the renormalized hadron
propagators in the $\sigma -\omega $ model. The numerical results are given
and discussed in Section 3. A summary is presented in Section 4.

\section{The models and coupled Dyson-Schwinger equations}

For the $\sigma -\omega $ model [10], the Lagrangian density is given by 
\begin{eqnarray}
\mathcal{L}_{\sigma -\omega } &=&-\overline{\psi }(\gamma _\mu k_\mu +M)\psi
-\tfrac 12(\partial _\mu \phi \partial _\mu \phi +m_{\mathrm{s}}^2\phi
^2)+g_{\mathrm{s}}\overline{\psi }\psi \phi  \nonumber \\
&&-\tfrac 14F_{\mu \nu }F_{\mu \nu }-\tfrac 12m_{\mathrm{v}}^2A_\mu A_\nu
-ig_{\mathrm{v}}\overline{\psi }\gamma _\mu \psi A_\mu +\mathcal{L}_{\mathrm{%
CTC}}  \tag{2}
\end{eqnarray}
where $F_{\mu \nu }=\partial _\mu A_\nu -\partial _\nu A_\mu $, $\partial
_\mu =\frac \partial {\partial x_\mu }$, $x_\mu =(\vec x,ix_0)$, $x^2=x_\mu
x_\mu =\vec x^2-x_0^2$ with $x_0\equiv t$, and the CTC means the counterterm
correction introduced for the purpose of renormalization. We shall use the
same notation as Refs. [8, 9]. The DS equations in the dressed HF scheme
(Fig. 1) can be written in the following form:

(a) for baryon 
\begin{equation}
G(k)=G^0(k)+G^0(k)\Sigma (k)G(k)=-[\gamma _\mu k_\mu -iM+\Sigma (k)]^{-1}, 
\tag{3a}
\end{equation}
\begin{equation}
\Sigma (k)=\Sigma _{\mathrm{s}}(k)+\Sigma _{\mathrm{v}}(k)  \tag{3b}
\end{equation}
\begin{equation}
\Sigma _{\mathrm{s}}(k)=-g_{\mathrm{s}}^2\int \frac{d^nq}{(2\pi )^4}%
\overline{G}(q)\overline{\Delta }_s(k-q)\Gamma _{\mathrm{s}}(k,q,k-q)+\Sigma
_{CTC}^{\mathrm{s}}(k),  \tag{4a}
\end{equation}
\begin{equation}
\Sigma _{\mathrm{v}}(k)=g_{\mathrm{v}}^2\int \frac{d^nq}{(2\pi )^4}\gamma
_\eta \overline{G}(q)\overline{D}_{\eta \lambda }(k-q)\Gamma _\lambda
(k,q,k-q)+\Sigma _{CTC}^{\mathrm{v}}(k),  \tag{4b}
\end{equation}

(b) for $\sigma $-meson 
\begin{equation}
\Delta _{\mathrm{s}}(k)=\Delta _{\mathrm{s}}^0(k)+\Delta _{\mathrm{s}%
}^0(k)\Pi _{\mathrm{s}}(k)\Delta _{\mathrm{s}}(k)=-i[k^2+m_{\mathrm{s}%
}^2+i\Pi _{\mathrm{s}}(k)-i\epsilon ]^{-1},  \tag{5a}
\end{equation}
\begin{equation}
\Pi _{\mathrm{s}}(k)=\chi g_{\mathrm{s}}^2\dint \frac{d^nq}{(2\pi )^4}%
\mathrm{Tr}[\overline{G}(k+q)\Gamma _{\mathrm{s}}(k+q,q,k)\overline{G}%
(q)]+\Pi _{CTC}^{\mathrm{s}}(k);  \tag{5b}
\end{equation}

(c) for $\omega $-meson 
\begin{equation}
D_{\mu \nu }(k)=D_{\mu \nu }^0(k)+D_{\mu \eta }^0(k)\Pi _{\eta \lambda
}(k)D_{\lambda \nu }(k),  \tag{6a}
\end{equation}
\begin{equation}
D_{\mu \nu }^0=(\delta _{\mu \nu }-\frac{k_\mu k_\nu }{k^2})\Delta _{\mathrm{%
v}}^0(k)  \tag{6b}
\end{equation}
\begin{equation}
\widehat{\Pi }_{\eta \lambda }(k)=-\chi g_{\mathrm{v}}^2\dint \frac{d^nq}{%
(2\pi )^4}\mathrm{Tr}[\gamma _\eta \overline{G}(k+q)\Gamma _\lambda (k+q,q,k)%
\overline{G}(q)].  \tag{6c}
\end{equation}
In the above equations $\Gamma _s$ and $\Gamma _\lambda $ (denoted by a
heavy dot in Fig. 1) are the $\sigma $-baryon and $\omega $-baryon vertex
functions, respectively; $n=4-\delta $ $(\delta \rightarrow 0^{+})$ and in
Eq. (3) the Feynman prescription $M\rightarrow M-i\varepsilon $ is
understood. In this paper, we shall only consider $\Gamma _{\mathrm{s}}=1$
and $\Gamma _\lambda =\gamma _\lambda $. We assume the following formula is
identical. 
\begin{equation}
\widehat{\Pi }_{\mu \nu }(k)=(\delta _{\mu \nu }-\frac{k_\mu k_\nu }{k^2})%
\widehat{\Pi }_{\mathrm{v}}(k),  \tag{6d}
\end{equation}
which implies $\widehat{\Pi }_{\mathrm{v}}(k)=\frac 13\sum_\mu \widehat{\Pi }%
_{\mu \mu }(k)$. From Eq. (6d) one observes that the renormalized $\Pi _{\mu
\nu }(k)$ can be obtained from 
\begin{equation}
\Pi _{\mathrm{v}}(k)=\widehat{\Pi }_{\mathrm{v}}(k)+\Pi _{\mathrm{CTC}}^{%
\mathrm{v}}(k).  \tag{6e}
\end{equation}
$\overline{G}$ ($\overline{\Delta }$, $\overline{D}$) denotes an appropriate
expression chosen for the calculation of the baryon (meson) propagator in
the self-energy. Just as in Ref. [9], we shall study the four schemes shown
in Table 1, where the first column gives the name of each scheme, while the
second and third explain how its $\Sigma _{\mathrm{s}}$ [$\Sigma _{\mathrm{v}%
}$] and $\Pi _{\mathrm{s}}$ [$\Pi _{\mathrm{v}}$] are obtained.

In the potential scheme P, $\Sigma ($P$)$ and $\Pi ($P$)$ are obtained by
setting $\overline{G}=-[\gamma _\mu k_\mu -iM_t]^{-1}$ where $M_t$ is the
true baryon mass and $\overline{\Delta }_{\mathrm{s}}=\Delta _{\mathrm{s}}^0 
$ ($\overline{D}=D^0$);

$\Sigma ($EP$)$ and $\Pi ($EP$)$ in the extended potential scheme EP are
obtained by setting $\overline{G}=G($P$)$ and $\overline{\Delta }_{\mathrm{s}%
}=\Delta _{\mathrm{s}}($P$)$ ($\overline{D}=D($P$)$);

To obtain $\Sigma ($BP$)$ and $\Pi ($BP$)$ for scheme BP, one sets $%
\overline{G}=G($BP$)$, $\overline{\Delta }_{\mathrm{s}}=\Delta _{\mathrm{s}%
}^0$ ($\overline{D}=D^0$), which implies that the baryon propagator has to
be determined self-consistently;

For $\Sigma ($FSC$)$ and $\Pi ($FSC$)$, one sets $\overline{G}=G($FSC$)$, $%
\overline{\Delta }_{\mathrm{s}}=\Delta _{\mathrm{s}}($FSC$)$ ($\overline{D}%
=D($FSC$)$), i.e. all the baryon and meson propagators are calculated
self-consistently.

It is known [3, 10] that in the zero density case, the baryon self-energy $%
\Sigma (k)=\gamma _\mu k_\mu a(k^2)-iMb(k^2)$. For convenience of
discussion, the case I (II) refers to neglecting (considering properly) the
contribution of $k_\mu k_\nu $ terms in $D_{\mu \nu }$. For the $\sigma
-\omega $ model we shall denote $a=a_{\mathrm{s}}+a_{\mathrm{v}}^\delta $
and $b=b_{\mathrm{s}}+b_{\mathrm{v}}^\delta $ in case I, while $a=a_{\mathrm{%
s}}+a_{\mathrm{v}}^\delta +a_{\mathrm{v}}^\Delta $ and $b=b_{\mathrm{s}}+b_{%
\mathrm{v}}^\delta +b_{\mathrm{v}}^\Delta $ in case II, where the
superscript '$\delta $' denotes the contribution of $\delta $-term in Eq.
(1a), and the '$\Delta $' is the contribution of $k_\mu k_\nu $-terms in Eq.
(1a). To fix the renormalization counterterms, we shall follow Ref. [8] and
use the on-shell renormalization condition on baryon and the intermediate
renormalization condition on mesons, which can be written as [9]: 
\begin{equation}
\Sigma _\eta (k)|_{\gamma _\mu k_\mu =iM_t}=0;\qquad \frac \partial
{\partial (\gamma _\mu k_\mu )}\Sigma _\eta (k)|_{\gamma _\mu k_\mu =iM_t}=0;
\tag{7a}
\end{equation}
\begin{equation}
\Pi _\eta (k)|_{k^2=0}=0;\qquad \frac \partial {\partial k^2}\Pi _\eta
(k)|_{k^2=0}=0,  \tag{7b}
\end{equation}
where $\eta =\mathrm{s}$, or $\mathrm{v}$. In addition, we shall also use ($%
\alpha (k^2)$, $\beta (k^2)$) and $\rho (k^2)$ to denote the baryon and
meson spectral weight functions, respectively. As is wellknown, one has 
\begin{equation}
\alpha (k^2)=\frac 1\pi \mathrm{Im}[\frac{1+a(k^2)}{D(k^2)}],  \tag{8a}
\end{equation}
\begin{equation}
\beta (k^2)=\frac 1\pi \mathrm{Im}[\frac{1+b(k^2)}{D(k^2)}],  \tag{8b}
\end{equation}
\begin{equation}
D(k^2)=[1+a(k^2)]^2k^2+[1+b(k^2)]^2M_t^2;  \tag{8c}
\end{equation}
\begin{equation}
\rho _\eta (k^2)=\frac 1\pi \mathrm{Im}[i\Delta _\eta (k^2)],\qquad \eta =%
\mathrm{s},\mathrm{v}  \tag{9}
\end{equation}
Using the renormalization conditions (Eqs. (7)), we can obtain the
expressions of the self-energy and the spectral weight functions. For the
nuclear matter as well as neutron matter, the explicit formulae of $\Sigma
_\eta (k^2)$ in $\sigma -\omega $ model are the same as in [8, 9]. while for
the $\Pi _\eta (k)$ in Eqs. (5, 6), $\chi =1$ is for the neutron matter and $%
\chi =2$ for the nuclear matter. So, Eqs. (2-6, 8, 9) and their explicit
expressions yield the closed set of the renormalized DS equations used for
our calculation.

\section{The numerical results}

We shall use the following values for the coupling constants and masses: 
\[
\begin{tabular}{ll}
$M_t=4.7585\mathrm{fm}^{-1},$ &  \\ 
$m_{\mathrm{s}}=2.6353\mathrm{fm}^{-1},$ & $m_{\mathrm{v}}=3.9680\mathrm{fm}%
^{-1}$ \\ 
$\overline{g}_{\mathrm{s}}^2=g_{\mathrm{s}}^2/16\pi ^2=0.5263,\;$ & $%
\overline{g}_{\mathrm{v}}^2=g_{\mathrm{v}}^2/8\pi ^2=1.3685$%
\end{tabular}
\]
The notation is the same as in Refs. [8, 9].

Now, we consider the $\sigma -\omega $ model in case II , where the
additional contribution of the $k_\mu k_\nu $-terms in Eq. (1a) is taken
into account. Following the method and rule obtained in [8], we solve the
coupled set of DS equations by the method of iteration. The numerical
results are shown in Figs. 2-4. In our calculation, we have studied three
different schemes: schemes P, EP and FSC. The baryon spectral functions
obtained from these three schemes are very close to each other. Though the
self-consistency makes the peak of the resonance higher, as a whole its
effect is not significant, just as found in case I [9]. However, comparing
Fig. 2 with Fig. 8 in [9], one observes there is a great quantitative change
in the functional behavior. The maxima of $\alpha (k^2)$ and $\beta (k^2)$
become more distinct and sharper in case II.

Let us designate $k^2>(<)-(M_t+m_{\mathrm{v}})^2$ as region I (II). We note $%
a_{\mathrm{v}}$ and $b_{\mathrm{v}}$ are real in region I and become complex
in region II, so in region II their imaginary parts will also contribute.
Let us fix $\overline{g}_{\mathrm{v}}^2=1.3685$ and consider, for instance,
the variation of $\alpha (k^2)$ with $\overline{g}_{\mathrm{s}}^2$ (see Fig.
3). When $\overline{g}_{\mathrm{s}}^2$ is small, there are two resonances
which are located in region I and II, respectively. Eq. (8a) can be rewriten
as $\alpha (k^2)=\frac 1\pi [\frac{a_{\mathrm{i}}D_{\mathrm{r}}-(1+a_{%
\mathrm{r}})D_{\mathrm{i}}}{D_{\mathrm{r}}^2+D_{\mathrm{i}}^2}]$, where the
subscript 'r' and 'i' denote the real and imaginary parts of $a(k^2)$ and $%
D(k^2)$, respectively. Comparing with ($a_{\mathrm{v}}$, $b_{\mathrm{v}}$),
the values of ($a_{\mathrm{s}}$, $b_{\mathrm{s}}$) is very small, and in
region I the contribution of ($a_{\mathrm{v}}$, $b_{\mathrm{v}}$) to the
denominator of $\alpha (k^2)$ is larger than to the numerator which is
mainly determined by the imaginary part of ($a_{\mathrm{s}}$, $b_{\mathrm{s}%
} $), so the resonance is small. If $\overline{g}_{\mathrm{s}}^2$ tends to
zero, the resonance in region II becomes more and more like the resonance of 
$\alpha (k^2)$ in the $\omega $ case and the resonance in region I
disappears (see [8]). As $\overline{g}_{\mathrm{s}}^2$ becomes larger, $a_{%
\mathrm{s}}$ and $b_{\mathrm{s}}$ are big. In this case the combined
contributions of ($a_{\mathrm{s}}$, $b_{\mathrm{s}}$) and ($a_{\mathrm{v}}$, 
$b_{\mathrm{v}}$) to the denominator and numerator of $\alpha (k^2)$ are
comparable in region I, thus the resonance becomes sharper. In region II
along with $|k^2|$ becoming larger $|D(k^2)|$ is big, so $\alpha (k^2)$
becomes small and the resonance in this region almost disappears, just as
shown in Fig. 2.

For both cases I and II, we have readjusted the parameters under the
condition that the spectral function $\alpha \left( k^2\right) $ should be
non-negative. We find that for the parameters of physical interest the
resonance is always in region II for case I, while the stronger one is in
region I for case II. Moreover, the resonance in case II is more distinct
than that in case I. Our results show the difference between these two cases
will remain even by the readjustment of the parameters. Thus, the
contribution of $k_\mu k_\nu $-term is very important to the baryon spectral
functions. Since $\alpha (k^2)$ relates directly to the probability of
occurrence of an excited baryon state, the contribution of $k_\mu k_\nu $%
-term in the $\omega $-meson propagator seems to make the possibility of
forming a resonance baryon state greater.

From Fig. 4, it looks that the effect of self-consistency on mesons is
discernible. However, there is no need to require self-consistency in the
meson propagators, because in Figs. 2 and 4 the results of scheme EP\ and
FSC are almost the same. Compared with Fig. 10 in [9] the self-consistent
meson spectral function is larger, but this change is not great. It means
the contribution of $k_\mu k_\nu $-terms to ($\rho _{\mathrm{s}}$, $\rho _{%
\mathrm{v}}$) is not important.

From Fig. 2, Fig. 3 in [9] and Fig. 2b in [8], one notes that there also
exists a cancellation between the effects on the self-consistency due to the 
$\sigma $ and $\omega $ mesons in case II.

In order to study the effect of self-consistency more carefully, we have
drawn $\alpha (k^2)$ and $\beta (k^2)$ calculated for an intermediate $%
\overline{g}_{\mathrm{v}}^2=0.3400$ in Fig. 5. One observes that the effect
of self-consistency is not significant, except for the region of the peaks,
where it makes the peak value a little higher. On the whole, we may say that
similarly to case I, the effects of self-consistency in case II are also not
important.

For the neutron matter the result are almost the same as above.

\section{Summary}

In this paper the coupled set of DS equations in the $\sigma -\omega $ for
two cases were solved self-consistently. The calculations show that in the $%
\sigma -\omega $ model, there is no need to require self-consistency in
meson propagators and the self-consistency almost has no effect on the
baryon propagator. Compared with case I, there is a distinct change in case
II in the baryon spectral functions which become more peaked. Such a change
cannot be removed even by a proper readjustment of parameters. Moreover,
there also exists a cancellation between the effects of the self-consistency
due to the $\sigma $ and $\omega $ mesons.

Our results show that in a calculation where the law of the baryon current
(BC) conservation is violated, the contribution of the $k_\mu k_\nu $terms
is generally not negligible and may serve as a sign signifying the degree of
its violence. If we assume the approximation made for the calculation is
appropriate, though BC may not be conserved, then the $k_\mu k_\nu $ terms
have to be considered (Way-$k_\mu k_\nu $). Clearly whether the results
obtained by Way-$k_\mu k_\nu $ are acceptable has still to be confirmed by a
more rigorous calculation where the laws incorporated in the model are
respected. Since a bare baryon-meson vertex is used, the RSCHF approximation
does not satisfy the Ward-Takahashi (W-T) identity. It has been pointed out
in Ref. [6] that this is the main reason why the BC conservation is
violated. Thus, if vertices consistent with the W-T identity are used
(Way-WT), the $k_\mu k_\nu $ terms may be neglected. Sofar no numerical
calculations along this line have been reported. However, we do agree with
Ref. [2] that this is a procedure worthy of pursuing, as it will also tell
whether and when the simpler Way-$k_\mu k_\nu $ or some other simple
approximation may be a good substitute for Way-WT, because the latter is
quite complicated except for some simple cases.

The work is supported in part by the National Natural Science Foundation of
China and the Foundation of Chinese Education Ministry.

\newpage\

\newpage\ 

\textbf{Table}

\begin{description}
\item[Table 1]  : Different calculation schemes.
\end{description}

\begin{tabular}{|l|l|l|}
\hline
Name & $\Sigma _{\mathrm{s}}(\Sigma _{\mathrm{v}})$ & $\Pi _{\mathrm{s}}(\Pi
_{\mathrm{v}})$ \\ \hline
P & $G_\Sigma ^0,\Delta _{\mathrm{s}}^0(D^0)$ & $G_\Sigma ^0$ \\ \hline
EP & $G($P$),\Delta _{\mathrm{s}}($P$)(D($P$))$ & $G($P$)$ \\ \hline
BP & $G($BP$),\Delta _{\mathrm{s}}^0(D^0)$ & $G($BP$)$ \\ \hline
FSC & $G($FSC$),\Delta _{\mathrm{s}}($FSC$)(D($FSC$))$ & $G($FSC$)$ \\ \hline
\end{tabular}

\newpage\ 

Figure captions

\begin{description}
\item[Fig. 1]  : Diagrammatic representation of the different
self-consistent (dressed) HF schemes. a. the baryon propagator; b. the $%
\omega $-meson propagator ; c. the $\sigma $-meson propagator.
\end{description}

\begin{description}
\item[Fig. 2]  : The baryon spectral functions $\alpha (k^2)$ and $\beta
(k^2)$ for the $\sigma -\omega $ model in case II.

\item[Fig. 3]  : The baryon spectral functions $\alpha (k^2)$ and $\beta
(k^2)$ for the $\sigma -\omega $ model in case II, left: ($\overline{g}_{%
\mathrm{s}}^2,\overline{g}_{\mathrm{v}}^2$)=(0.0500, 1.3685); right: ($%
\overline{g}_{\mathrm{s}}^2,\overline{g}_{\mathrm{v}}^2$)=(0.0100, 1.3685).
\end{description}

\begin{description}
\item[Fig. 4]  : The meson spectral functions $\rho _\lambda (k^2)$ for the $%
\sigma -\omega $ model in case II. top: $\sigma $-meson: $\lambda =\mathrm{s}
$; bottom: $\omega $-meson: $\lambda =\mathrm{v}$.
\end{description}

\begin{description}
\item[Fig. 5]  :The baryon spectral functions $\alpha (k^2)$ and $\beta (k^2)
$ for the $\sigma -\omega $ model in case II, ($\overline{g}_{\mathrm{s}}^2,%
\overline{g}_{\mathrm{v}}^2$)=(0.5263, 0.3400).
\end{description}

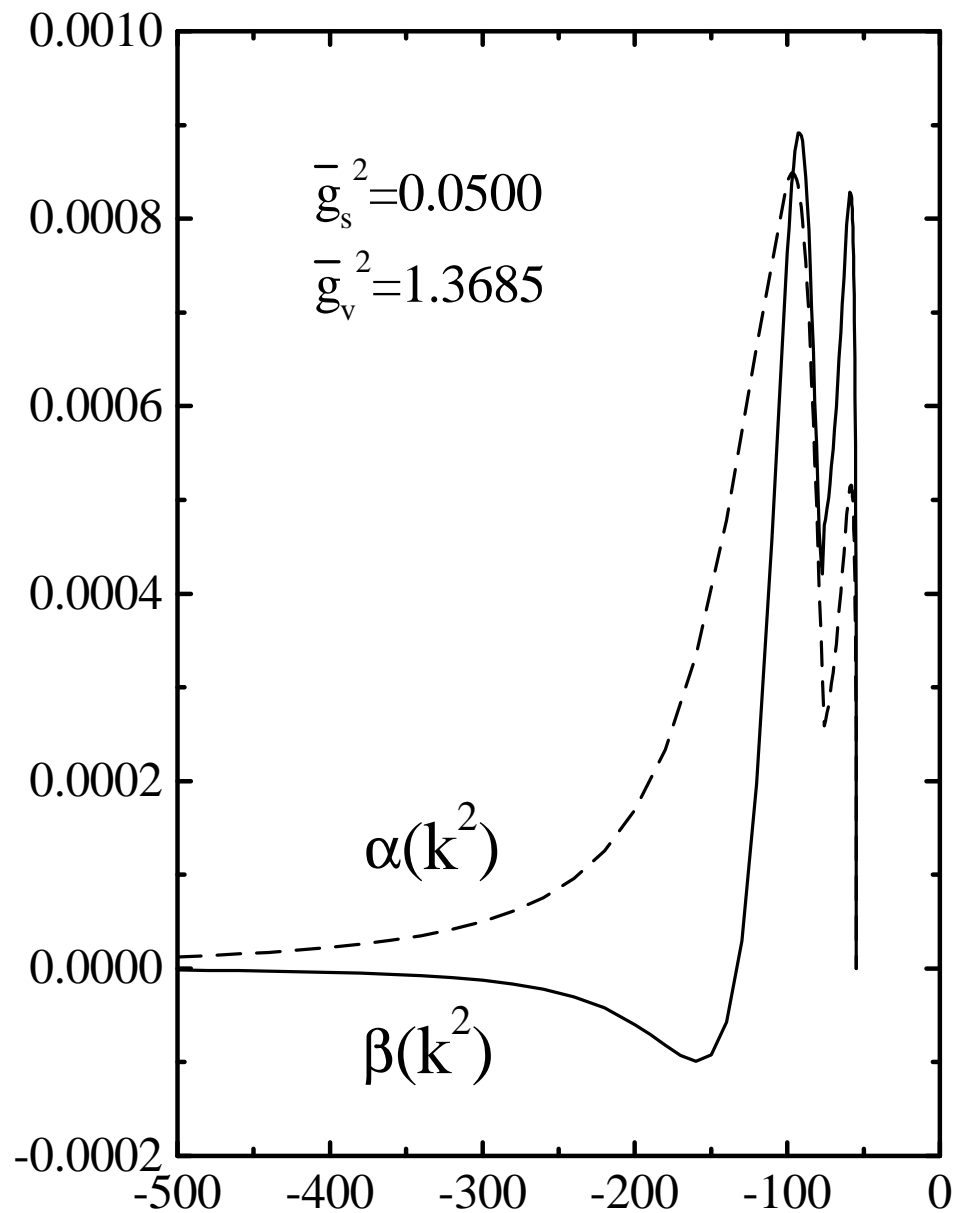
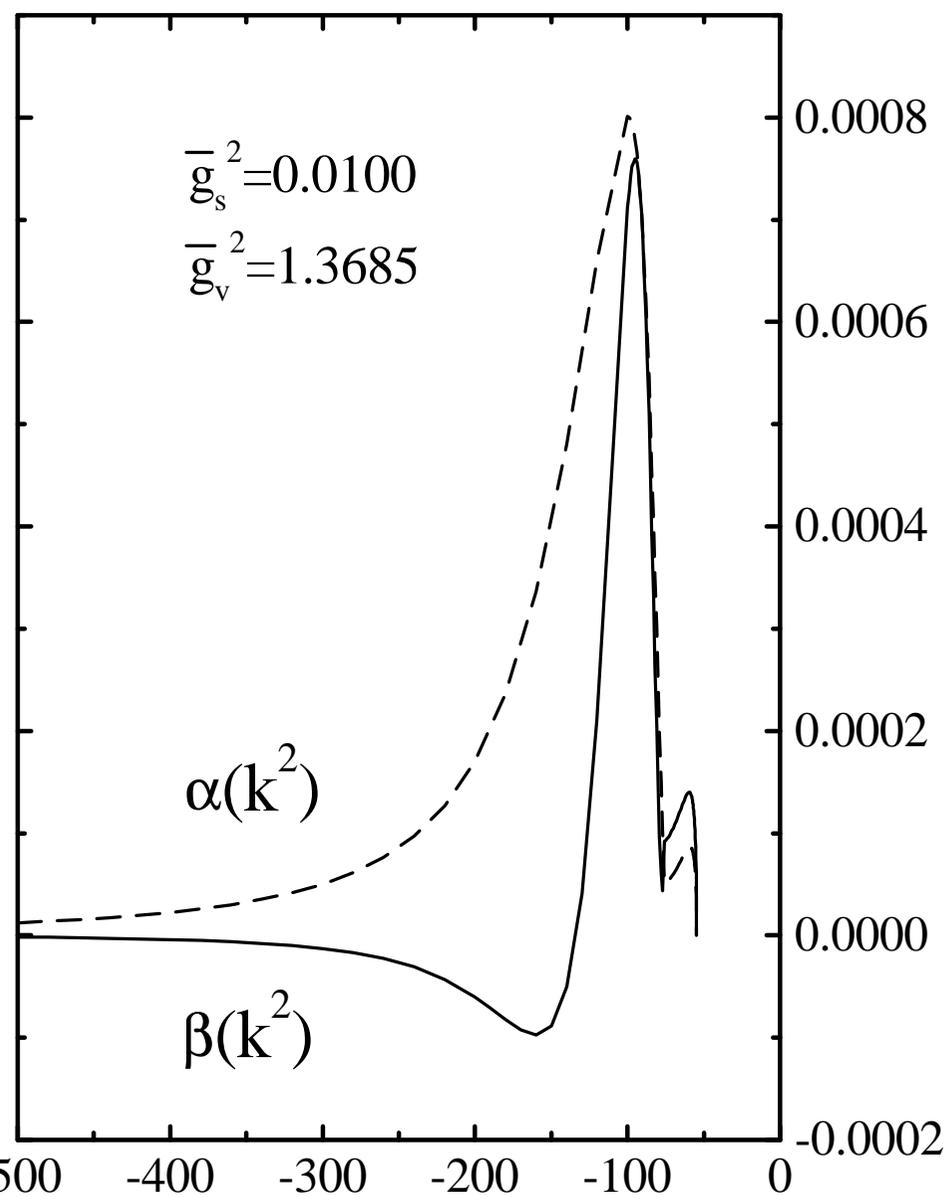
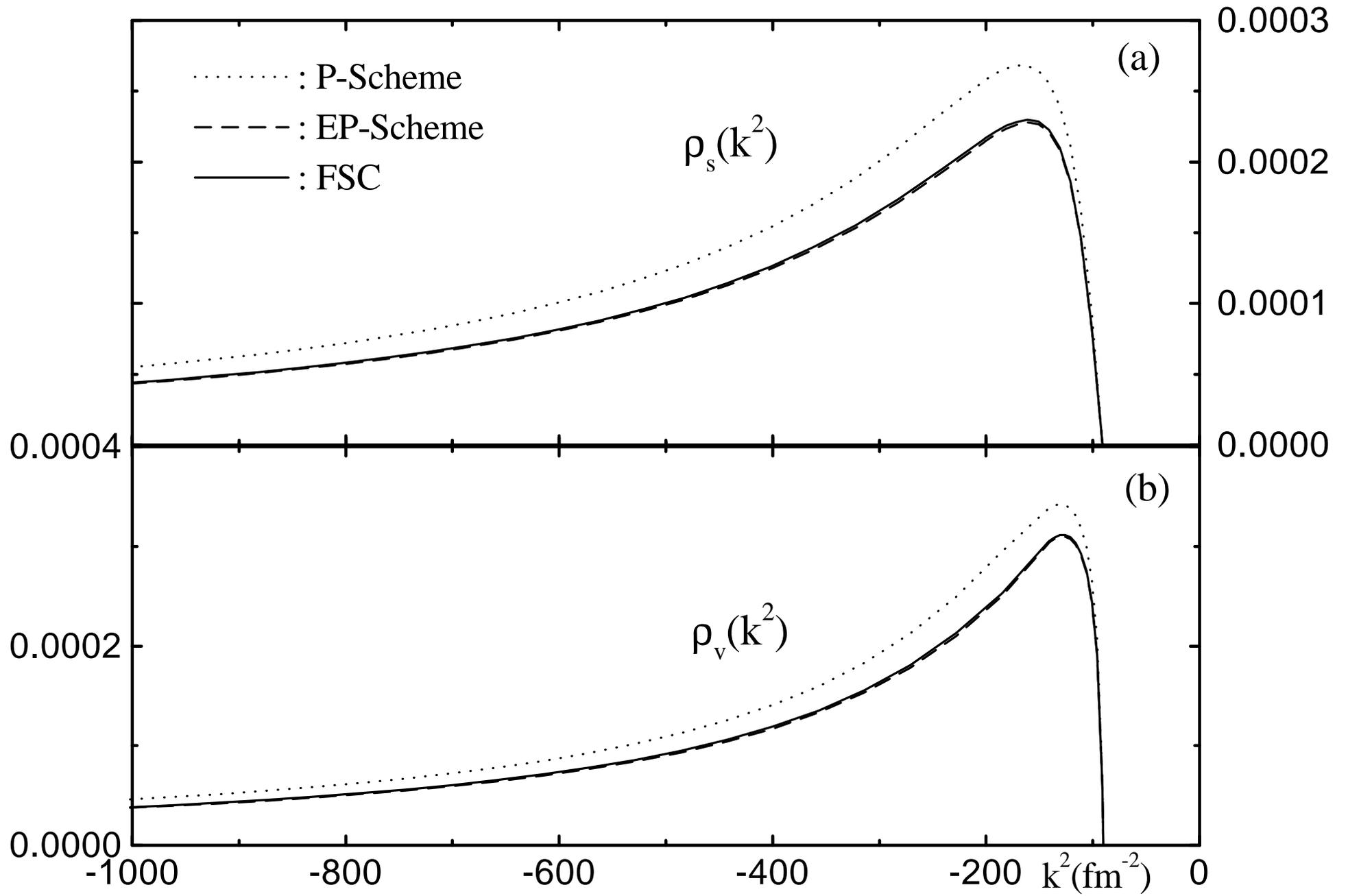
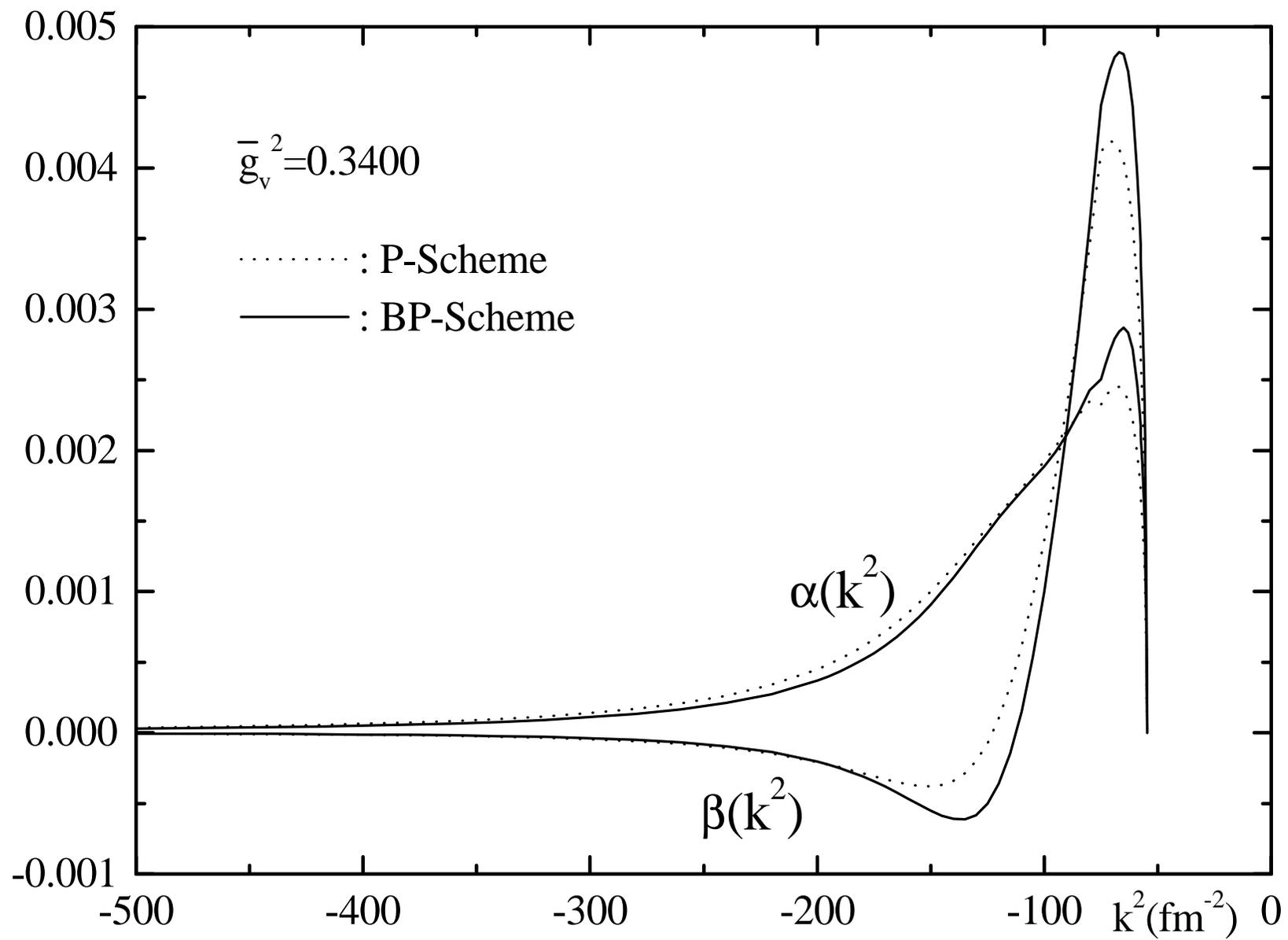
\end{document}